\documentclass[a4paper]{article}

\usepackage[english]{babel}
\usepackage[utf8x]{inputenc}
\usepackage[T1]{fontenc}
\usepackage{comment}

\usepackage[a4paper,top=3cm,bottom=2cm,left=3cm,right=3cm,marginparwidth=1.75cm]{geometry}

\usepackage{amsmath,amssymb}
\usepackage{graphicx}
\usepackage[colorinlistoftodos]{todonotes}
\usepackage[colorlinks=true, allcolors=blue]{hyperref}
\usepackage{authblk}
\newcommand\blue[1]{{\color{black}#1}}
\newcommand\red[1]{{\color{black}#1}}

\title{Ultracompact minihalos associated with \\stellar-mass primordial black holes}
\author[1,2]{Tomohiro Nakama}
\affil[1]{{\small Department of Physics and Astronomy, The Johns Hopkins University,

3400 N. Charles St., Baltimore, MD 21218, USA}}
\affil[2]{{\small Institute for Advanced Study, The Hong Kong University of Science and Technology, 

Clear Water Bay, Kowloon, Hong Kong}}

\author[3,4,5]{Kazunori Kohri}
\affil[3]{{\small Theory Center, IPNS, KEK, 1-1 Oho, Tsukuba, Ibaraki 305-0801, Japan}}
\affil[4]{{\small The Graduate University for Advanced Studies (SOKENDAI), 1-1 Oho, Tsukuba, Ibaraki 305-0801, Japan}}
\affil[5]{{\small Rudolf Peierls Centre for Theoretical Physics, University of Oxford, Department of Physics, 

Oxford, OX1 3PU, UK}}

\author[1,6,7]{Joseph Silk}
\affil[6]{{\small Institut d'Astrophysique, UMR 7095 CNRS, Sorbonne Universit\'{e}, 98bis Blvd Arago, 75014 Paris, France}}
\affil[7]{{\small Beecroft Institute of Particle Astrophysics and Cosmology, Department of Physics, 

University of Oxford, Oxford OX1 3RH}}

\date{}

\begin{document}
\maketitle
\begin{abstract}
The possibility that primordial black hole binary mergers  of stellar mass can  explain the signals detected by the gravitational-wave interferometers has attracted much attention. In this scenario, primordial black holes can comprise only part of the entire dark matter, say, of order   0.1 \%.  This implies that most of the dark matter is accounted for by a different component, such as Weakly Interacting Massive Particles. We point out that in this situation, very compact dark matter minihalos, composed of the dominant component of the dark matter, are likely to be formed abundantly in the early Universe, with their formation redshift and abundance depending on primordial non-Gaussianity. They may be detected in future experiments via pulsar observations. 
\end{abstract}
\section{Introduction}
Binary mergers of primordial black holes (PBHs) \cite{Hawking:1971ei,novikov,Carr:1974nx}\footnote{See Refs. \cite{Carr:2009jm,Carr:2016drx} for observational limits on PBHs on various masses and see Refs. \cite{Nakama:2013ica,Nakama:2014fra} for numerical simulations of PBH formation.} of stellar mass have been recently proposed as an  explanation for  
the gravitational waves \cite{Abbott:2016blz} detected by the gravitational-wave interferometers 
 \cite{Bird:2016dcv,Sasaki:2016jop,Clesse:2016vqa} (see also Ref.~\cite{Sasaki:2018dmp} for an overview). In this scenario, primordial black holes may comprise only part of the entire dark matter, say, around 0.1 \% \cite{Sasaki:2016jop}. This implies that the most of the dark matter is accounted for by a different component, such as Weakly Interacting Massive Particles (WIMPs) \cite{Bertone:2004pz}. Among various mechanisms for PBH formation \cite{Carr:2005zd}, the collapse of primordial fluctuations, enhanced on small scales relative to large-scale fluctuations, is most often discussed in the literature, and we focus here on this mechanism.
  In order to produce a sufficient number of PBHs, in most cases, the power spectrum or the root-mean-square amplitude of primordial fluctuations has to be enhanced considerably on scales corresponding to the masses of PBHs under consideration relative to the amplitude of primordial fluctuations on large scales \cite{Carr:1975qj}, as determined by the experiments measuring cosmic microwave background (CMB) radiation \cite{Aghanim:2018eyx}. This enhancement leads to potentially observable  CMB spectral distortions \cite{Chluba:2012we,Kohri:2014lza,Nakama:2017xvq} or a stochastic gravitational wave  background \cite{Saito:2008jc,Saito:2009jt,Inomata:2016rbd,Nakama:2016gzw,Kohri:2018awv,Byrnes:2018txb,Inomata:2018epa,Cai:2018dig}, in addition to the formation of PBHs. This enhancement also leads to the formation of compact dark matter minihalos at redshifts substantially larger than those for standard structure formation \cite{Kohri:2014lza}. How much primordial power needs to be enhanced on small scales depends on primordial non-Gaussianity, when the abundance of PBHs is fixed \cite{Byrnes:2012yx,Nakama:2016kfq,Nakama:2016gzw,Nakama:2017xvq}. In this paper, we show that ultracompact minihalos (UCMHs), which can be detectable by future pulsar-timing experiments, can be formed  abundantly provided that  we assume that PBHs account for those black holes whose existence has been revealed by the gravitational-wave interferometers. We also show how this conclusion depends on primordial non-Gaussianity. See also Ref. \cite{Adamek:2019gns} for a related study that focuses on thermal freeze-out  dark matter. We note however that the increasingly sensitive direction detection limits on WIMPs are focusing attention on dark matter particles that may not have a detectable annihilation or scattering signature, and our discussion below applies to generic (cold) dark matter particles.

\section{UCMHs associated with stellar-mass PBHs}
If the stellar-mass PBH formation probability is relatively large, with such stellar-mass PBHs being a sub-dominant component of the dark matter, then UCMHs comprised of the dominant component of the dark matter such as WIMPs are expected to be formed abundantly \cite{Kohri:2014lza}, unless matter fluctuations are not erased by free streaming or interactions on the corresponding scales \cite{Bringmann:2009vf}. However, when and how many UCMHs are formed depend on primordial non-Gaussianity \cite{Nakama:2016kfq,Nakama:2016gzw,Nakama:2017xvq}, as shown below. 

First, the mass of a PBH is roughly given by the mass of radiation within the comoving length of the region collapsing to a PBH at the moment when this scale reenters the horizon. On the other hand, the mass of the dark matter within the same comoving scale is smaller by a factor $a/a_{\mathrm{eq}}$, assuming PBHs are formed at the scale factor $a$. Hence, the mass scale $M$ of UCMHs associated with PBHs with mass $M_H$ is
\begin{equation}
M\sim \frac{a}{a_{\mathrm{eq}}}M_H=\left(\frac{M_H}{M_{\mathrm{eq}}}\right)^{1/2}M_H\sim 6\times 10^{-8}M_\odot\left(\frac{M_H}{10M_\odot}\right)^{3/2},
\end{equation}
noting $M_H\sim t\sim a^2$. Note that the horizon mass at \red{equality} appearing here is $M_{\mathrm{eq}}\simeq 2.8\times 10^{17}M_\odot$ \cite{Nakama:2016gzw}. The damping scale of WIMPs is model dependent \cite{Bringmann:2009vf} and can be smaller than the above  UCMH mass. The free-streaming scale is also very small for Planckian interacting dark matter \cite{Garny:2015sjg}. This argument for UCMH formation associated with stellar-mass PBHs would also apply to PBHs much lighter than the typical UCMH mass above, say $10^{-12}M_\odot$, being the dominant component of the dark matter, as considered in Ref. \cite{Inomata:2017vxo}. The free-streaming scale of PBHs is at most on the order of the Hubble radius at their formation, and hence the effect of free streaming on non-linear structures that formed later is practically zero.

Let us assume a delta-function-type spectrum of the curvature perturbation \red{leading} to PBH formation:
\begin{equation}
{\cal P}={\cal A}^2k\delta(k-k_*).
\end{equation}
The root-mean-square amplitude of density perturbations is \cite{Nakama:2017qac}
\begin{equation}
\sigma(z)=\frac{2{\cal A}D_1(a)}{5\Omega_{m0}H_0^2}k_*^2T(k_*),
\end{equation}
where $D_1(a)=a$ during the matter domination and for $k\gg k_{\mathrm{eq}}$ the transfer function is \cite{Dodelson:2003ft}
\begin{equation}
T(k)=\frac{12k_{\mathrm{eq}}^2}{k^2}\ln\frac{k}{8k_{\mathrm{eq}}}. 
\end{equation}
The above formula reflects the logarithmic growth during  radiation domination after \red{horizon} reentry, and also the growth in proportion to the scale factor during \red{matter} domination. Let us introduce $z_*$ by $\sigma(z_*)=1$. At around this redshift, all the dark matter in the Universe would collapse to UCMHs. The corresponding scale factor $a_*$ is written as 
\begin{equation}
a_*=\frac{5\Omega_{m0}H_0^2}{24{\cal A}k_{\mathrm{eq}}^2}\left(\ln\frac{k_*}{8k_{\mathrm{eq}}}\right)^{-1}.
\end{equation}

UCMH formation for a fixed PBH formation probability depends on the statistics of primordial perturbations. In Refs. \cite{Nakama:2016kfq,Nakama:2016gzw,Nakama:2017xvq}, the following phenomenological non-Gaussianity model was used:
\begin{equation}
P(\zeta)=\frac{1}{2\sqrt{2}\tilde{\sigma}\Gamma(1+1/p)}\exp\left[-\left(\frac{|\zeta|}{\sqrt{2}\tilde{\sigma}}\right)^p\right].
\end{equation}
This reduced to a Gaussian PDF when $p=2$. 
If we fix the parameter $p$ and the PBH abundance $\beta$, the amplitude $\cal A$ is determined, then the redshift of UCMH formation $z_*$ is also correspondingly determined.  

The amplitude $\cal A$ is determined by the non-Gaussian parameter $p$ and the PBH abundance $\beta$ by
\begin{equation}
{\cal A}=\left[\frac{2\Gamma(1+3/p)}{3\Gamma(1+1/p)}\right]^{1/2}\frac{2^{-1/2}\zeta_c}{[Q^{-1}(1/p,2\beta)]^{1/p}}.
\end{equation}
We choose the threshold of PBH formation $\zeta_c=0.67$. See Ref. \cite{Nakama:2017xvq} for a discussion about this choice of threshold. We set $\beta=10^{-11}$, which corresponds to $f=\Omega_{\mathrm{PBH}}/\Omega_{\mathrm{DM}}\sim 0.001$ at a PBH mass of $10M_\odot$ \cite{Nakama:2016gzw}. The redshift $z_*$ as a function of $p$ is shown in Fig. 1. 
\begin{figure}[htbp]
  \begin{center}
    \includegraphics[clip,width=7.0cm]{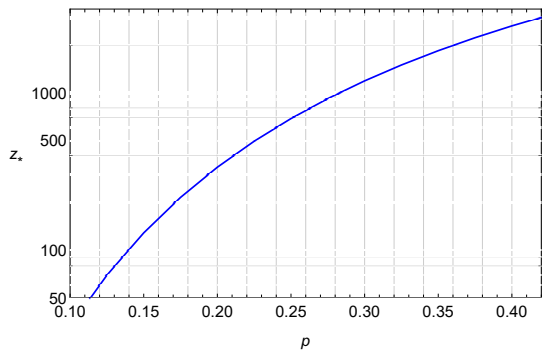}
    \caption{The UCMH formation redshift $z_*$ as a function of the non-Gaussian parameter $p$, assuming stellar-mass PBHs comprise 0.1\% of the dark matter, whereas the mass of dark matter particles of the dominant component is much smaller than $\sim 10^{-7}M_\odot$. }
    \label{fig:hamu}
  \end{center}
\end{figure}
Thus, UCMH formation is rather generic, and for $p>0.4$, UCMHs are formed shortly after equality. Note that before \red{equality} dark matter overdensities $\delta$ can collapse when they are locally matter-dominant ($ \delta\rho_m\sim\rho_R$), that is $\delta\sim a_{\mathrm{eq}}/a$, and hence UCMH formation well before \red{equality} is suppressed.  UCMH formation is avoided when \red{extreme} non-Gaussianity is realized, such as in  the model of \cite{Nakama:2016kfq} or for even smaller values of $p$. One can repeat the above calculations for other types of non-Gaussianity, such as local-type, quadratic or cubic non-Gaussianity, as in Refs. \cite{Nakama:2016gzw,Nakama:2017xvq}. However, the above $p$-type non-Gaussianity is most generic, in the sense that a wider range of $\sigma$ can be realized for a fixed abundance of PBHs. See Fig. 5 of Ref. \cite{Nakama:2017xvq}. The values of $p$ which give the same value of $\sigma$ for each $f_{\mathrm{NL}}$, setting $\beta=10^{-11}$, are shown in Fig. \ref{pvsfnl}. 
\begin{figure}[htbp]
  \begin{center}
    \includegraphics[clip,width=7.0cm]{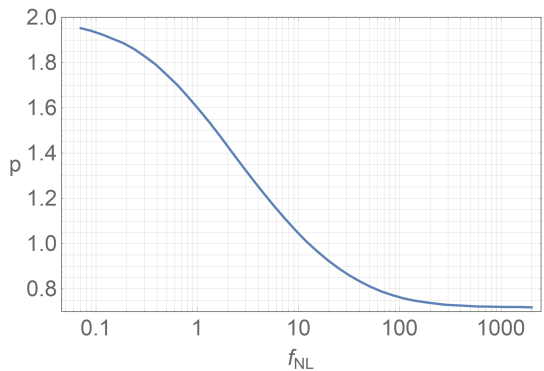}
    \caption{The values of $p$ which give the same value of $\sigma$ for each $f_{\mathrm{NL}}$, setting $\beta=10^{-11}$. We find $p\rightarrow 0.72$ for $f_{\mathrm{NL}}\rightarrow \infty$.}
    \label{pvsfnl}
  \end{center}
\end{figure}

After UCMHs are formed, they would lose part of their mass due to tidal effects during large-scale structure formation and also within our galaxy, and this mass loss might have important implications for the detectability of UCMHs. First, let us note that the dark matter, which may or may not be in the form of UCMHs,  \blue{would accumulate around PBHs. In a homogeneous, matter-dominated Universe, the mass of a dark matter halo surrounding a PBH grows in proportion to the scale factor \cite{Ricotti:2009bs}, so most of the dark matter would be bound in structures surrounding PBHs at redshift of order unity, if the fraction of PBHs in the dark matter is 0.001 and in addition if other kinds of inhomogeneities were absent. This shows that the enhancement of structure formation due to the presence of PBHs is negligible, since standard structure formation, seeded by standard nearly-scale-invariant primordial fluctuations, takes place at redshifts of order ten. During this epoch, most UCMHs would become part of larger-scale structures.\footnote{Structure formation on scales larger than $k_*^{-1}$ may also take place after UCMHs are formed, but in order to fully clarify this issue, one would need cosmological simulations assuming a sharp spike in the primordial spectrum, as we have assumed in this paper. For Gaussian fluctuations one may also apply the standard analytical method such as Press–Schechter formalism \cite{Press:1973iz} to quantify this effect, but this formalism is based on the density perturbation smoothed over different scales, larger than $k_*^{-1}$ in the current context. Mostly, window functions such as real-space top-hat, Gaussian or sharp-k filter \cite{Ando:2018qdb}, are often used, but in the case of the delta-function spectrum, structure formation on scales larger than $k_*^{-1}$ is highly sensitive to the choice of the window function. For instance, it is highly suppressed if we use the sharp-$k$ filter or Gaussian window function \cite{Kohri:2014lza,Nakama:2017qac}. In this case, UCMHs, abundantly formed at high redshifts as a result of the collapse of primordial perturbations with comoving wave number $k_*$ would simply become part of standard larger-scale halos before experiencing mergers among UCMHs. On the other hand, if we use the real-space top-hat window function, structure formation seeded by the delta-function spectrum on scales larger than $k_*^{-1}$ would be more efficient. If initially-formed UCMHs become part of larger halos whose formation time and mass are relatively close to the formation times and original masses of UCMHs, then the formed halos themselves may be regarded as UCMHs, which may also survive and be detectable by pulsar timing. On the other hand, some UCMHs would become part of halos seeded by the delta-function spectrum whose formation time is significantly later than that of UCMHs and whose mass is significantly larger than UCMHs, in this case, UCMHs would retain their original mass, and later on, if their host becomes part of our galaxy, the host may experience significant mass loss, and UCMHs may get liberated, to be observed by pulsar timing. }} 
After UCMHs become part of larger halos, in the region where our galaxy is formed, their hosts would experience disk shocking to lose ${\cal O}(10)$\% of \red{their} mass at a distance $\simeq 8.5$kpc from the Galactic center, as shown later. That is, a significant fraction of UCMHs would be liberated from their hosts. Since these liberated UCMHs \blue{had been} gravitationally bound by their host halos only relatively loosely, \red{their} tidal stripping due to the global tides of these host halos would not be so significant. However, these liberated UCMHs would further experience disk shocking in our galaxy, which we estimate as follows. As shown later, the energy per unit mass gained by particles comprising UCMHs as a result of one disk crossing is $\Delta E\sim GRr^2\Sigma_d(R)^2/M(R)$, where $r$ is the radius of UCMHs, $R=8.5$kpc is the distance from the Galactic center and $\Sigma_d(R)$ is the disk surface mass density at $R$ and $M(R)$ is the total mass of our galaxy enclosed within $R$. The magnitude of the gravitational potential $\Phi$ of UCMHs would be $\Phi\sim Gm/r$, where $m$ is the mass of UCMHs. Denoting the number of disk crossings by $N$, the ratio $N\Delta E/\Phi$ would measure effectiveness of disk shocking. It can be written as $\sim R\Sigma_d^2(R)/[M(R)\rho_{m}(a_*)]\sim 0.2(a_*/0.01)^3$, setting $N=100$. Hence, disk shocking of UCMHs would not be so significant. Note that even if UCMHs lose some fraction of their mass, \red{they are} still detectable by pulsar timing as long as \red{their} mass is larger than the detectable threshold mass by pulsar timing, \red{which can be significantly smaller than the original UCMH mass of $\sim 10^{26}$g that we consider}. On the other hand, the hosts of UCMHs are formed at lower redshifts, and hence their tidal stripping by disk shocking is more effective, and they lose ${\cal O}(10)$\% of their mass, as discussed more in detail in the next section. 

A PBH fraction of order $10^{-2}$ on the mass scale $10^{26}$g should be probed in future by pulsar timing \cite{Kashiyama:2012qz}, due to additional acceleration of the observed pulsars from close encounters with PBHs, and this equally implies that  UCMHs associated with stellar-mass PBHs would also be probed in the future by pulsar timing. 
The UCMH radius is $a_*/k_*\sim 1000$AU, for $k_*=10^6\mathrm{Mpc}^{-1}$ and $a_*=10^{-3}$. This is comparable to the minimum impact parameter for pulsar timing, so the assumption of point \red{masses} adopted in Ref. \cite{Kashiyama:2012qz} to discuss PBHs would be justifiable also in the context of UCMHs. \blue{Probing minihalos formed from the standard nearly-scale-invariant primordial fluctuations by pulsar timing is discussed in Ref. \cite{Kashiyama:2018gsh}.}

\section{Disk shocking of the host halos of UCMHs}
Let us present the Navarro–Frenk–White (NFW) profile and a few of its properties following Ref. \cite{Lokas:2000mu}:
\begin{equation}
\frac{\rho(r)}{\rho_c(z)}=\frac{\delta}{(r/r_s)(1+r/r_s)^2}.
\end{equation}
We define the virial radius as the radius within which the mean density is $v=178$ times the critical density $\rho_c(z)$ (see Ref. \cite{Lokas:2000mu} and references therein). The characteristic density $\delta$ is given in terms of the concentration $c=r_v/r_s$ as 
\begin{equation}
\delta=\frac{vc^3g(c)}{3},\quad g(c)=\frac{1}{\ln (1+c)-c/(1+c)}. 
\end{equation}
The mass within the virial radius is $M_v=(4/3)\pi r_v^3v\rho_c(z)$. The mass enclosed within $s=r/r_v$ is \cite{Lokas:2000mu}
\begin{equation}
M(s)=4\pi\delta\rho_c(z)r_s^3\int_0^x\frac{x^2dx}{x(1+x)^2}=g(c)\left[\ln(1+cs)-\frac{cs}{1+cs}\right]M_v,
\end{equation}
noting $\int ^x xdx/(1+x)^2=\ln (1+x)-x/(1+x)$. $M(s=1)=M_v$, as it should \red{be}. 
This behaves as $M(s)\simeq g(c)M_v r^2/(2r_s^2)$ for $c s\ll 1$. 
The gravitational potential is 
\begin{align}
\Phi(s)=-\int_r^{\infty}\frac{GM(r)}{r^2}dr
&=g(c)GM_vr_s^{-2}\left[\int_x^\infty\frac{\ln(1+x)}{x^2}dx-\int_x^{\infty}\frac{1}{x(1+x)}dx\right]\nonumber\\
&=-g(c)V_v^2\frac{\ln(1+cs)}{s},\quad V_v^2=\frac{4}{3}\pi G r_v^2 v\rho_c(z).
\end{align}
Note that $\Phi(s)\simeq-(c-c^2s/2)g(c)V_v^2$ for $cs\ll1$.
Once a halo becomes part of a larger halo, the evolution of halos would be mainly determined by tidal stripping, instead of accretion. In this case, the halo profile may be well described by the NFW profile before becoming part of a larger halo, truncated at some radius $r_t$ due to tidal stripping experienced after becoming part of a larger halo. 

The non-dimensional matter power spectrum for a Harrison-Zel'dovich-Peebles spectrum is \cite{Dodelson:2003ft}
\begin{equation}
\Delta^2(k)=\frac{k^3P(k)}{2\pi^2}=\delta_H^2\left(\frac{k}{H_0}\right)^4T^2(k)\left(\frac{D_1(a)}{D_{1,0}}\right)^2,
\end{equation}
where $\delta_H\sim 4.6\times 10^{-5}$, $T(k)=12k_{\mathrm{eq}}^2/k^2\ln(k/k_{\mathrm{eq}})$ for $k\gg k_{\mathrm{eq}}\simeq 0.01 \mathrm{Mpc}^{-1}$, $D_1(a)=a$ during  matter domination, and $D_{1,0}\simeq 0.8$. Let us consider some comoving scale $r\sim \pi/k$, which reaches turnaround at redshift $z$ with $R_{\mathrm{turn}}\simeq ar_0/2$ \cite{Lyth:2009zz}. The eventual virial radius is $r_v\sim R_{\mathrm{turn}}/2$, hence we have $k=\pi a/4r_v$. For the wavenumber corresponding to $r_v=10\mathrm{kpc}$, the amplitude becomes $\Delta(k)\simeq 1$ at $z\simeq 6$. That is, minihalos orbiting around $10\mathrm{kpc}$ from the Galactic center became part of a non-linear region at $z\simeq 6$, assuming that region was typical, at which growth due to accretion from the background Universe halted. After this moment, the evolution of such minihalos would be characterized by tidal stripping. 

Hence we consider minihalos described by the above NFW profile at $z=6$ as an example, and we estimate the truncation radius determined by disk shocking due to the Galactic disk as follows. 
Note that the collapsed fraction of the Universe at $z=6$ is $\sim 60\%$ according to an ellipsoidal collapse model employed in Ref. \cite{Angulo:2009hf}.

The change $\Delta v_z$ of the particle velocity relative to the minihalo in one passage through the Galactic disk at distance $R$ from the Galactic center is \cite{1972ApJ...176L..51O}
\begin{equation}
\Delta v_z\simeq \frac{2zg_m(R)}{V(R)},
\end{equation}
where $z$ is the position of the particle relative to the center of the minihalo, $V(R)=\sqrt{GM(R)/R}$ is the velocity of the minihalo and $g_m$ is the maximum acceleration due to the disk. $g_m$ is related to the disk surface density at a distance $R$ from the Galactic center via \cite{1995ApJ...438..702K}
\begin{equation}
g_m(R)=2\pi G\Sigma_d(R).
\end{equation}
We use $\Sigma_d(R)=\Sigma_0\exp(-R/R_d)$, with $(\Sigma_0,R_d)=(753 M_\odot \mathrm{pc}^{-2},3 \mathrm{kpc}),(182 M_\odot \mathrm{pc}^{-2},3.5 \mathrm{kpc})$ for the thin and thick disk, respectively \cite{McMillan:2011wd}. Note that $V(R)$ should be determined by the bulge mass and the dark matter mass, as well as the disk mass. We assume that the bulge mass is $8.9\times 10^9M_\odot$ \cite{McMillan:2011wd}, and also that the dark matter profile of the Milky Way galaxy is 
\begin{equation}
\rho(r)=\frac{0.0125 M_\odot \mathrm{pc}^{-3}}{(r/r_s)(1+r/r_s)^2}, \quad r_s=17 \mathrm{kpc}.
\end{equation}
We set $R$ to the solar radius $R_\odot$ of $8.5$ kpc \cite{McMillan:2011wd}. 
The energy gain per unit mass is 
\begin{equation}
\Delta E(R,r)\simeq\frac{2r^2g_m^2(R)}{3V^2(R)},
\end{equation}
where $z^2$ has been replaced by $(1/2)\int d\cos \theta r^2 \cos^2\theta=r^2/3$. For particles which rotate sufficiently fast around the minihalo center, the above energy gain would be suppressed. Hence we multiply the above $\Delta E$ by the adiabatic correction factor $A(x)=(1+x^2)^{-3/2}$, with the adiabatic parameter $x=\omega(r)\tau(R)$ \cite{Gnedin:1999rg}. The orbital frequency is $\omega(r)=v(r)/r=\sqrt{Gm(r)/r^3}$, and $\tau(R)=H/V(R)$ is the disk crossing time with $H$ denoting the half-thickness of the disk, for which we choose $H=100$ pc. For $c s\ll1$ as well as $x\gg 1$, 
\begin{equation}
A(x)\simeq \left(\frac{g(c)GM_v}{2r_s^2r}\right)^{-3/2}\tau^{-3}(R).
\end{equation}
The number of disk \red{crossings} over a time period of $T_{\mathrm{MW}}=10$ Gyr is
\begin{equation}
N_{\mathrm{cross}}(R)=\frac{V(R)T_{\mathrm{MW}}}{\pi R}.
\end{equation}
Particles at $r$, relative to the minihalo's center, which orbits at distance $R$ from the Galactic center, would gain an energy of order $N_{\mathrm{cross}}(R)\Delta E(R,r)$. If it exceeds the absolute magnitude of the gravitational potential of the minihalo at $r$, then such particles would leave the minihalo. 
 Hence we assume the minihalo is truncated at $r_t$, which satisfies \begin{equation}
N_{\mathrm{cross}}(R)\Delta E(R,r_t)A(x)=-\Phi(s_t=r_t/r_v).
\end{equation}
Note that virial radius $r_v$ here is that specified at redshift $z=6$, at which the minihalo has not experienced significant tidal stripping. As a result, the minihalo's mass becomes $m(r_t)$. 

The above equation can be rewritten as 
\begin{equation}
s_t^3A[x(s_t)]=c_1g(c)\ln (1+cs_t), \quad c_1=\frac{2\pi v V^2(R)G\rho_c(z)}{N_{\mathrm{cross}}(R)g_m^2(R)}\simeq 0.028,
\end{equation}
and also
\begin{equation}
\omega(r)\tau(R)=c_2\left(\frac{g(c)}{s^3}\left[\ln(1+cs)-\frac{cs}{1+cs}\right]\right)^{1/2},\quad c_2=\left(\frac{4\pi vG \rho_c(z)}{3}\right)^{1/2}\tau(R)\simeq 1.5\times 10^{-3}.
\end{equation}
Hence, $s_t$ does not depend on the initial mass of the  minihalo. 
It turns out that the dependence of $s_t\simeq c_1^{1/3}$ on concentration is also weak and for $c>10$, $s_t\simeq c_1^{1/3}\simeq 0.3$. Let us introduce
\begin{equation}
\eta(c)\equiv\frac{M(s_t)}{M_v}=\frac{\ln (1+cc_1^{1/3})-cc_1^{1/3}/(1+c c_1^{1/3})}{\ln(1+c)-c/(1+c)}.
\end{equation}
We find $\eta=0.43,0.68,0.8,1$ for $c=10,100,1000,\infty$. Standard halos have concentration of 10 or larger with logarithmic dependence on mass and hence these halos would have lost ${\cal O}(10)$ \% of their masses according to this estimate, which is also consistent with the order-of-magnitude argument in the previous section. To conclude, ${\cal O}(10)$ \% of UCMHs have likely been liberated from their hosts at distance 8.5 kpc from the Galactic center, and their mass would not be significantly smaller than their initial mass at formation, as discussed in the previous section. Hence, they would probably be detectable by future pulsar-timing experiments. 
\section{Discussion}
The level of non-Gaussianity needed to avoid \blue{expected} UCMH limits discussed here is even stronger than that required to avoid current/future pulsar timing limits on induced gravitational waves associated with PBH formation \cite{Nakama:2016gzw}. Our conclusion is that UCMHs associated with PBHs are likely to survive until today without experiencing substantial mass loss, whereas in Ref. \cite{Nakama:2017qac} we conservatively neglected minihalos which are formed as a result of enhanced small-scale primordial power and which become part of larger standard halos, when deriving upper limits on primordial power on small scales by gamma rays or neutrinos from those minihalos. In that work, the enhancement of primordial power was less substantial than that in this paper, which implies later formation redshifts and shallower gravitational potential wells of formed minihalos, so the mass loss of these minihalos during hierarchical structure formation would be correspondingly more important. 

Recently, PBHs with masses $10^{-11}M_\odot<M<10^{-6}M_\odot$ have been constrained in Ref. \cite{Niikura:2017zjd}. One may wonder whether compact dark matter halos can also be constrained by microlensing. The Einstein radius for a point mass $M$ is 
\begin{equation}
R_E^2=\frac{4GM}{c^2}D,\quad D\equiv D_L^2\left(\frac{1}{D_L}-\frac{1}{D_S}\right).
\end{equation}
Let us consider a point mass with $10^{-6}M_\odot$ and $D=100$ kpc as an example. The Einstein radius is $\sim \sqrt{r_sD}\sim 10^6$ km, and a clump with $\rho\sim M/R_E^3\sim 10^{-22}M_\odot \mathrm{km}^{-3}$ would yield a lensing signal similar to a point mass. The current critical density is $10^{11}M_\odot \mathrm{Mpc}^{-3}\sim10^{-37}M_\odot \mathrm{km}^{-3}$, and hence objects which collapsed as early as \red{matter-radiation} equality would not reach such a high density. Excitingly, M31b may have an earth mass PBH, at a 1\% level in terms of possible dark matter fraction \cite{Niikura:2019kqi}. 
The corresponding UCMHs are probably too small to be detectable by pulsar timing, since the detectable smallest mass of PBHs was shown to be around $10^{22}$g in Ref. \cite{Kashiyama:2012qz}, but they could affect the Kuiper Belt \cite{Penarrubia:2019wei}.

Compact objects might also be probed by astrometry \cite{Garcia-Bellido:2017fdg}. In order for a point mass with $M_\odot$ to cause velocity change $\Delta v$ of order $1 \mathrm{km}/s$ of an observed star, the impact parameter $b$ has to be $\sim 10^{11}\mathrm{m}$ \cite{Feldmann:2013hqa}. The dark matter density near the Sun is $\sim 0.01M_\odot \mathrm{pc}^{-3}$. Let us assume compact objects of $M_\odot$, such as primordial black holes, account for the entire dark matter. Then the number density $n$ of such objects is $\sim 0.01\mathrm{pc}^{-3}$. The probability of a star acquiring $\Delta v\sim 1 \mathrm{km}/s$ per second due to a close encounter with such an object is $b^2vn\sim 10^{-23}s^{-1}$, with the relative velocity $v=100 \mathrm{km}/s$. If a billion stars are observed for a year, the event rate is $10^{-7}$, so probing small compact objects with astrometry would probably be challenging. \blue{This rate is in proportion to the mass of compact objects, so probing smaller objects is even more difficult. }

Future astrometry experiments such as Small-Jasmine may also measure proper motions of stars in a nearby dark-matter-dominated dwarf galaxy such as Sculptor \cite{Massari}. Typical velocities $v$ of stars are $10\mathrm{km/s}$, and we assume these are related to the mass of the total dark matter $M$ inside a radius of $R=100$pc via \red{$v^2=GM/R$}. A star would acquire a velocity change $\Delta v$ on the order of $Gm/bv$ due to a close encounter with a compact object with  impact parameter $b$, where $m=10^{26}\mathrm{g}$ is the mass of the compact object. For $\Delta v=1\mathrm{km}/\mathrm{s}$, $b\sim 10^6\mathrm{m}$, and this velocity change arises over a time-scale of $100\mathrm{s}$. The event rate $\Gamma$ is written as \red{$Gmv/R^2(\Delta v)^2$}, which is estimated to be \red{$\Gamma\sim 10^{-26}\mathrm{s}^{-1}$}. We will further consider such dynamical signatures in a follow-up paper.

We have mentioned additional acceleration of observed pulsars caused by close encounters with UCMHs, associated with the existence of stellar-mass PBHs, but there is another effect, which is the Shapiro time delay. This latter effect probes larger-mass PBHs or dark matter halos \cite{Dror:2019twh}. See also Ref. \cite{Clark:2015sha}.
In addition to gravitational or dynamical signals associated with UCMHs, there could be further signals from these UCMHs, if the dark matter annihilation is sufficiently efficient. For instance, they can give local ionization hot spots that could be an ionization source in molecular clouds \cite{Silk:2018orr}.

\section*{Acknowledgments}
We thank Keisuke Inomata for a helpful discussion. T.N. thanks KEK for
hospitality received during this work. T.N. was partially supported by
JSPS Overseas Research Fellowships. K.K. was partially supported by JSPS
KAKENHI Grants No.~JP17H01131, No. 26247042, and MEXT KAKENHI
Grants No.~JP15H05889, No.~JP16H0877, No. JP18H04594, and No. JP19H05114.


\end{document}